\documentclass[twocolumn,showpacs,pra,aps,floatfix]{revtex4}

\usepackage{graphicx}
\usepackage{bm}
\usepackage{psfrag}
\usepackage{amsmath}
\usepackage{amssymb}
\usepackage{latexsym}
\usepackage{exscale}

\begin{document}

\title{Local-field correction to the spontaneous decay rate of atoms\\
embedded in bodies of finite size}

\author{Ho Trung Dung}
\affiliation{Institute of Physics, Academy of
Sciences and Technology, 1 Mac Dinh Chi Street,
District 1, Ho Chi Minh city, Vietnam}

\author{Stefan Yoshi Buhmann}

\author{Dirk-Gunnar Welsch}

\affiliation{Theoretisch-Physikalisches Institut,
Friedrich-Schiller-Universit\"at Jena,
Max-Wien-Platz 1, D-07743 Jena, Germany}

\date{\today}

\begin{abstract}
The influence of the size and shape of a dispersing and absorbing
dielectric body on the local-field corrected spontaneous-decay of an
excited atom embedded in the body is studied on the basis of the
real-cavity model. By means of a Born expansion of the Green tensor
of the system it is shown that to linear order in the susceptibility
of the body the decay rate exactly follows Toma\v{s}'s formula found
for the special case of an atom at the center of a homogeneous
dielectric sphere [Phys.\ Rev.\ A \textbf{63}, 053811 (2001)]. It is
further shown that for an atom situated at the interior of an
arbitrary dielectric body this formula remains valid beyond the
linear order. The case of an atom embedded in a weakly polarizable
sphere is discussed in detail.
\end{abstract}

\maketitle


\section{Introduction}
\label{sec_intro}

It has long been recognized that when an atom is situated in a
material medium, the local electromagnetic field acting on it differs
from the macroscopic one due to the gaps between the atom and the
surrounding medium atoms, which are ignored on a coarse-grained
macroscopic scale \cite{Lorentz1880,Onsager36}. Accounting for the
difference between the two fields hence requires a correction---the
local-field correction. Classical calculations of local-field effects,
which typically have their origin in near dipole-dipole interactions,
can be found in textbooks (see, e.g., Ref.~\cite{Lorentz52}). In
quantum theory, investigations of local-field effects are often
related to the problem of the spontaneous decay of an excited
guest atom embedded in a (dielectric) host. Local-field effects in
spontaneous decay have been studied, e.g., for crystals
\cite{Knoester80,Vries98} and disordered dielectrics
\cite{Juzeliunas97,Fleischhauer99,Crenshaw00,Berman04} on the basis on
microscopic models for coupled atomic dipoles.

In macroscopic descriptions, local-field effects are frequently taken
into account by regarding the guest atom as being enclosed in a
virtual \cite{Lorentz1880,Scheel99a} or real (spherical) cavity
\cite{Onsager36,Glauber91,Scheel99b,Tomas01,Rahmani02,Ho03}
surrounded by the medium, with the cavity size being small compared to
the relevant transition wavelength. The cavity in the former model is
virtual in the sense that it does not perturb the macroscopic field.
It is filled by the atoms comprising the medium, which produce no net
effect at the central position in important special cases such as
cubic or random structures. In the latter model, the field is modified
by the presence of the cavity, which is an empty region containing
only the guest atom. Microscopic models often tend to agree with the
virtual-cavity results
\cite{Knoester80,Juzeliunas97,Fleischhauer99,Berman04},
while many recent experiments on spontaneous emission in dielectrics
support the real-cavity model
\cite{Rikken95,Lavallard96,Schuurmans98,Kumar03}. It has been presumed
that while the virtual-cavity model applies to interstitial atoms, the
real-cavity model is specific to substitutional atoms, and that the
case of substitutional atoms occurs prevalently for impurity atoms in
disordered dielectrics~\cite{Vries98}.

Since in the approaches to the local-field effects, the host medium
has typically been assumed to be a bulk medium that extends homogeneously
to infinity, the question of the effect of the size and shape of the
host medium on the local field has arisen. In a macroscopic approach,
the spontaneous-decay rate of an excited atom in some free-space
region can be given in terms of the imaginary part of the Green tensor
of the macroscopic Maxwell equations, which characterizes the
(macroscopic) environment of the atom. This relation in principle
allows for including local-field corrections for atoms embedded in
arbitrary material configurations by assuming a real (spherical)
cavity surrounding the atom and calculating the corresponding Green
tensor. Using the real-cavity model, Toma\v{s} \cite{Tomas01} studied
the local-field correction to the spontaneous-decay rate of an excited
atom, which is located at the center of a dispersing and absorbing
dielectric sphere. Reformulating the result by representing it in a
form, which does not explicitly refer to the highly symmetric system
considered, he made the conjecture that it may also remain valid
beyond the specific example and hence also apply to other locations of
the atom and other shapes of the host body.

Based on a numerical computation of the respective Green tensors,
Rahmani and Bryant \cite{Rahmani02} considered the case of an atom at
an arbitrary location within a dielectric sphere or a dielectric cube.
Comparing their results for the dielectric sphere including the
local-field correction with earlier results disregarding the
local-field correction \cite{Chew88,Kim88}, they suggested a rate
formula, which in the case of weakly absorbing material corresponds to
Toma\v{s}'s formula. However, since their approach relies heavily on
numerical calculations, it cannot produce explicit expressions for the
quantities that are related to the local-field correction.

The exact analytical evaluations of the Green tensors of realistic
systems which have finite sizes and include a cavity can be very
cumbersome. In this paper, we present an attempt to overcome this
difficulty by writing the Green tensor as a Born series in terms of
the susceptibility, where in many situations one can restrict oneself
to several leading-order terms. In particular, we show that to linear
order the spontaneous-decay rate of a guest atom in a dielectric host
body can be separated into a term representing the local-field
correction to the decay rate in free space and a term related to the
scattering Green tensor of the body without the atom---a result which
exactly corresponds to Toma\v{s}'s conjecture mentioned above.
Furthermore, we show that for atoms which are situated at the interior
of a macroscopic body Toma\v{s}'s conjecture remains valid beyond the
linear order. We illustrate the theory by discussing in detail the
case of an atom embedded in a spherical dielectric body.

The paper is organized as follows. The basic equations for the
spontaneous-decay rate and the Born expansion of the Green tensor
determining the rate are given in Sec.~\ref{sec_BE}. They are
used in Sec.~\ref{subsec_rc} to study the problem of the local-field
corrected decay rate within the frame of the real-cavity model, and a
proof of Toma\v{s}'s formula is given. The examples of an atom
embedded in a bulk dielectric medium or in a dielectric sphere are
examined in Sec.~\ref{sec_ex}, followed by a summary (Sec.~\ref{sum}).


\section{Spontaneous-decay rate}
\label{sec_BE}

Consider an excited two-level electric-dipole emitter, henceforth
referred to as an atom, which is positioned at $\mathbf{r}_\mathrm{A}$
and surrounded by dispersing and absorbing dielectric bodies. The
spontaneous-decay rate can be given in the form of
\cite{Agarwal75,Ho00}
\begin{equation}
\label{e3}
\Gamma=\frac{2k_\mathrm{A}^2}{\hbar\varepsilon_0}\,
 \mathbf{d}_\mathrm{A}\!\cdot\!\mathrm{Im}\,\bm{G}
 (\mathbf{r}_\mathrm{A},\mathbf{r}_\mathrm{A},\omega_\mathrm{A})
 \!\cdot\!\mathbf{d}_\mathrm{A},
\end{equation}
where $\mathbf{d}_\mathrm{A}$ and $\omega_\mathrm{A}$ are the
(real) dipole matrix element and (shifted) frequency of the relevant
atomic transition, respectively, and $k_\mathrm{A}$ $\!=$
$\!\omega_\mathrm{A}/c$. The Green tensor of the bodies,
$\bm{G}(\mathbf{r},\mathbf{r}',\omega)$, satisfies the equation
\begin{align}
\label{e4}
&\hat{H}\,\bm{G}(\mathbf{r},\mathbf{r}',\omega)
 =\delta(\mathbf{r}-\mathbf{r}')\bm{I},
 \\
\label{e5}
&\hat{H}\equiv
 \bm{\nabla}\times\bm{\nabla}\times
 -\frac{\omega^2}{c^2}\,\varepsilon(\mathbf{r},\omega)
\end{align}
($\bm{I}$, unit tensor) together with the boundary condition
\begin{equation}
\label{e4b}
\bm{G}(\mathbf{r},\mathbf{r}',\omega)\to 0
\quad\mbox{for }|\mathbf{r}-\mathbf{r}'|\to \infty,
\end{equation}
where $\varepsilon(\mathbf{r},\omega)$ is the frequency- and
space-dependent complex permittivity which satisfies the
Kramers--Kronig relations. Note that satisfaction of the boundary
condit\-ion~(\ref{e4b}) is ensured by assuming
$\mathrm{Im}\,\varepsilon(\mathbf{r},\omega)$ $\!>$ $\!0$.

Equation (\ref{e3}) always applies when the atom is placed in
some free-space region. Separating the Green tensor into bulk and
scattering parts $\bm{G}^{(0)}$ and $\bm{G}^{(1)}$, respectively,
\begin{equation}
\label{e14.2}
\bm{G}(\mathbf{r},\mathbf{r}',\omega)
 =\bm{G}^{(0)}(\mathbf{r},\mathbf{r}',\omega)
 +\bm{G}^{(1)}(\mathbf{r},\mathbf{r}',\omega),
\end{equation}
and taking into account that in the case where the bulk part refers
to free space, the relation
\begin{equation}
\label{Eq.6}
\mathrm{Im}\,\bm{G}^{(0)}
 (\mathbf{r}_\mathrm{A},\mathbf{r}_\mathrm{A},\omega_\mathrm{A})
 =\frac{k_\mathrm{A}}{6\pi}\,\bm{I}
\end{equation}
holds (see, e.g., Ref.~\cite{Knoell01}), we may rewrite Eq.~(\ref{e3})
as
\begin{equation}
\label{e14.4}
\Gamma=\Gamma_0+\frac{2k_\mathrm{A}^2}{\hbar\varepsilon_0}\,
 \mathbf{d}_\mathrm{A}\!\cdot\!\mathrm{Im}\,\bm{G}^{(1)}
 (\mathbf{r}_\mathrm{A},\mathbf{r}_\mathrm{A},\omega_\mathrm{A})
 \!\cdot\!\mathbf{d}_\mathrm{A},
\end{equation}
where
\begin{equation}
\label{e14.5}
\Gamma_0=\frac{k_\mathrm{A}^3d_\mathrm{A}^2}
 {3\pi\hbar\varepsilon_0}
\end{equation}
is the spontaneous-decay rate in free space.

If the atom is embedded in a body, application of Eq.~(\ref{e3})
requires special care in two respects. Firstly, the coincidence limit
of the bulk part of the Green tensor diverges when the permittivity of
the body is complex, as is the case in general. Only if material
absorption can be neglected so that the permittivity can be regarded
as being real, $\varepsilon(\mathbf{r}_\mathrm{A},\omega_\mathrm{A})$
$\!\simeq$
$\!\mathrm{Re}\,\varepsilon(\mathbf{r}_\mathrm{A},\omega_\mathrm{A})$,
this limit exists,
\begin{equation}
\label{Eq.9}
\mathrm{Im}\,\bm{G}^{(0)}
 (\mathbf{r}_\mathrm{A},\mathbf{r}_\mathrm{A},\omega_\mathrm{A})
 =\sqrt{\varepsilon}\,\frac{k_\mathrm{A}}{6\pi}\,\bm{I}
\end{equation}
(see, e.g., Ref.~\cite{Knoell01}), leading to
\begin{equation}
\label{Eq.10}
\Gamma=\sqrt{\varepsilon}\,\Gamma_0
 +\frac{2k_\mathrm{A}^2}{\hbar\varepsilon_0}\,
 \mathbf{d}_\mathrm{A}\!\cdot\!\mathrm{Im}\,\bm{G}^{(1)}
 (\mathbf{r}_\mathrm{A},\mathbf{r}_\mathrm{A},\omega_\mathrm{A})
 \!\cdot\!\mathbf{d}_\mathrm{A}.
\end{equation}
Secondly, the Green tensor of the macroscopic Maxwell equations does
not account for the fact that the local field felt by the atom is
different from the macroscopic one in general. That is, even if
absorption is neglected, the rate formula (\ref{Eq.10}) is not
complete because it lacks the local-field corrections.


\subsection{Born expansion}
\label{sec_BE-1}

To calculate the (scattering part of the) Green tensor for an
arbitrary arrangement of dielectric bodies, it may be helpful
to use an appropriate Born expansion. Decomposing the permittivity as
\begin{equation}
\label{e6}
\varepsilon(\mathbf{r},\omega)
 =\overline{\varepsilon}(\mathbf{r},\omega)
 +\chi(\mathbf{r},\omega) ,
\end{equation}
and assuming that the solution
$\overline{\bm{G}}(\mathbf{r},\mathbf{r}',\omega)$ to the equation
\begin{equation}
\label{e7}
\hat{\overline{H}}\,
 \overline{\bm{G}}(\mathbf{r},\mathbf{r}',\omega)
 =\delta(\mathbf{r}-\mathbf{r}')\bm{I}
\end{equation}
is known [where $\hat{\overline{H}}$ is defined as in Eq.~(\ref{e5})
with $\overline{\varepsilon}$ instead of $\varepsilon$], the Green
tensor can be written in the form of a Born series,
\begin{align}
\label{e8}
&\bm{G}(\mathbf{r},\mathbf{r}',\omega)
 =\overline{\bm{G}}(\mathbf{r},\mathbf{r}',\omega)
 +\sum_{k=1}^\infty\Delta_k\bm{G}(\mathbf{r},\mathbf{r}',\omega),\\
\label{e9}
&\Delta_k\bm{G}(\mathbf{r},\mathbf{r}',\omega)
 =\Bigl(\frac{\omega}{c}\Bigr)^{2k}
 \Biggl[\prod_{j=1}^k\int\mathrm{d}^3s_j\,
 \chi(\mathbf{s}_j,\omega)\Biggr]\nonumber\\
&\quad\times
 \overline{\bm{G}}(\mathbf{r},\mathbf{s}_1,\omega)\!\cdot\!
 \overline{\bm{G}}(\mathbf{s}_1,\mathbf{s}_2,\omega)\!\cdots\!
 \overline{\bm{G}}(\mathbf{s}_k,\mathbf{r}',\omega),
\end{align}
as can be verified using the relationships
\begin{align}
\label{e10}
\hat{H}\,\Delta_1\bm{G}(\mathbf{r},\mathbf{r}',\omega)
=&\;\frac{\omega^2}{c^2}\chi(\mathbf{r},\omega)
 \bigl[\overline{\bm{G}}(\mathbf{r},\mathbf{r}',\omega)
\nonumber\\&\;
- \Delta_1\bm{G}(\mathbf{r},\mathbf{r}',\omega)\bigr],\\
\label{e11}
\hat{H}\,\Delta_k\bm{G}(\mathbf{r},\mathbf{r}',\omega)
=&\;\frac{\omega^2}{c^2}\chi(\mathbf{r},\omega)
 \bigl[\Delta_{k-1}{\bm{G}}(\mathbf{r},\mathbf{r}',\omega)
 \nonumber\\
&\; -\Delta_k\bm{G}(\mathbf{r},\mathbf{r}',\omega)\bigr]
 \qquad\mbox{for }k>1.
\end{align}

The expansion (\ref{e8}) of the Green tensor is valid for arbitrarily
spatially varying $\overline{\varepsilon}(\mathbf{r},\omega)$ and
$\chi(\mathbf{r},\omega)$. Obviously, it may be very useful when
$\chi$ can be regarded as being a (small) perturbation to
$\overline{\varepsilon}$ such that one makes only a small error by
disregarding the higher-order terms. In particular, this is the case
if the bodies are weakly polarizable, as we shall assume in the
following.


\subsection{Weakly polarizable bodies}
\label{subsec_wd}

For weakly polarizable bodies, it is natural to regard the
susceptibilities of the bodies as a small perturbation to the
free-space permittivity $\overline{\varepsilon}(\mathbf{r},\omega)$
$\!=$ $1$ $\!+$ $\!i\eta$ \mbox{($\eta$ $\!\to$ $\!0_+$)}, i.e.,
$|\chi(\mathbf{r},\omega)|$ $\!\ll$ $\!1$. This means that we focus on
frequencies that are sufficiently far from a resonance frequency of
the dielectric material. Note that the small (positive) imaginary part
of $\overline{\varepsilon}$ ensures that $\overline{\bm{G}}$ fulfills
the boundary condition according to Eq.~(\ref{e4b}) so that the
spatial integrals in Eq.~(\ref{e9}) converge. We have (see, e.g.,
Ref.~\cite{Knoell01})
\begin{equation}
\label{e13}
\overline{\bm{G}}(\mathbf{r},\mathbf{r}',\omega)
 =-\frac{\delta(\mathbf{u})}{3k^2}\bm{I}
 +\frac{k}{4\pi}(a\bm{I}
 -b\tilde{\mathbf{u}}\tilde{\mathbf{u}})
 e^{i\sqrt{1+i\eta}\,q}
,
\end{equation}
where
\begin{equation}
\label{e14}
a = a(q)=\frac{1}{q}+\frac{i}{q^2}-\frac{1}{q^3},
\quad
b = b(q)=\frac{1}{q}+\frac{3i}{q^2}-\frac{3}{q^3},
\end{equation}
$\mathbf{u}$ $\!=$ $\!\mathbf{r}$ $\!-$ $\!\mathbf{r}'$,
$\tilde{\mathbf{u}}$ $\!=$ $\!\mathbf{u}/u$, $k$ $\!=$ $\!\omega/c$,
$q$ $\!=$ $\!ku$.

Separating the Green tensor into bulk and scattering parts
in accordance with Eq.~(\ref{e14.2}), assuming the atom to be located
in a free-space region such that
\begin{equation}
\label{e14.3}
\mathrm{Im}\,\bm{G}^{(0)}
 (\mathbf{r}_\mathrm{A},\mathbf{r}_\mathrm{A},\omega_\mathrm{A})
 =\mathrm{Im}\,\overline{\bm{G}}
 (\mathbf{r}_\mathrm{A},\mathbf{r}_\mathrm{A},\omega_\mathrm{A})
 =\frac{k_\mathrm{A}}{6\pi}\,\bm{I}
\end{equation}
[cf.~Eq.~(\ref{Eq.6})], and applying Eq.~(\ref{e8}), we may represent
the scattering part of the Green tensor in the rate
formula~(\ref{e14.4}) as
\begin{equation}
\label{e14.6}
\bm{G}^{(1)}
 (\mathbf{r}_\mathrm{A},\mathbf{r}_\mathrm{A},\omega_\mathrm{A})
 =\sum_{k=1}^\infty\Delta_k\bm{G}
 (\mathbf{r}_\mathrm{A},\mathbf{r}_\mathrm{A},\omega_\mathrm{A}).
\end{equation}
For $\mathbf{r}$ and $\mathbf{r}'$ belonging to a free-space region,
substitution of Eq.~(\ref{e13}) into Eq.~(\ref{e9}) yields the first-
and second-order terms in the Born expansion, $\Delta_1\bm{G}$ and
$\Delta_2\bm{G}$, respectively, as follows:
\begin{align}
\label{e14.1}
&\Delta_1\bm{G}(\mathbf{r},\mathbf{r}',\omega)
 =\frac{k^4}{16\pi^2}\int\mathrm{d}^3s\,\chi(\mathbf{s},\omega)
 \bigl[aa'\bm{I}-ab'\tilde{\mathbf{u}}'\tilde{\mathbf{u}}'
 \nonumber\\
 &\quad-a'b\tilde{\mathbf{u}}\tilde{\mathbf{u}}
 +bb'(\tilde{\mathbf{u}}\!\cdot\!\tilde{\mathbf{u}}')
 \tilde{\mathbf{u}}\tilde{\mathbf{u}}'\bigr]
 e^{i\sqrt{1+i\eta}\,(q+q')}
\end{align}
[$\mathbf{u}$ $\!=$ $\!\mathbf{r}$ $\!-$ $\!\mathbf{s}$,
$q$ $\!=$ $\!ku$, $a$ $\!=$ $\!a(q)$, $b$ $\!=$ $\!b(q)$;
$\mathbf{u}'$ $\!=$ $\!\mathbf{s}$ $\!-$ $\!\mathbf{r}'$,
$q'$ $\!=$ $\!ku'$, $a'$ $\!=$ $\!a(q')$, $b'$ $\!=$ $\!b(q')$],
\begin{align}
\label{e16}
&\Delta_2\bm{G}(\mathbf{r},\mathbf{r}',\omega)=
 -\frac{k^4}{48\pi^2}\int\mathrm{d}^3s\,\chi^2(\mathbf{s},\omega)
\bigl[aa'\bm{I}-ab'\tilde{\mathbf{u}}'\tilde{\mathbf{u}}'
\nonumber\\
&\qquad -a'b\tilde{\mathbf{u}}\tilde{\mathbf{u}}
 +bb'(\tilde{\mathbf{u}}\!\cdot\!\tilde{\mathbf{u}}')
 \tilde{\mathbf{u}}\tilde{\mathbf{u}}'\bigr]
 e^{i\sqrt{1+i\eta}\,(q+q')}\nonumber\\
&+\frac{k^7}{64\pi^3}\!\int\!\mathrm{d}^3s_1\,
 \chi(\mathbf{s}_1,\omega)\!
 \int\!\mathrm{d}^3s_2\,\chi(\mathbf{s}_2,\omega)
 e^{i\sqrt{1+i\eta}\,(q_1+q_{12}+q_2)}
 \nonumber\\
&\quad\times\bigl[a_1a_{12}a_2\bm{I}
 -a_1a_{12}b_2\tilde{\mathbf{u}}_2\tilde{\mathbf{u}}_2
 -a_1b_{12}a_2\tilde{\mathbf{u}}_{12}\tilde{\mathbf{u}}_{12}
 \nonumber\\
&\qquad -b_1a_{12}a_2\tilde{\mathbf{u}}_1\tilde{\bm{u}}_1
 +a_1b_{12}b_2(\tilde{\mathbf{u}}_{12}
 \!\cdot\!\tilde{\mathbf{u}}_2)
 \tilde{\mathbf{u}}_{12}\tilde{\mathbf{u}}_2\nonumber\\
&\qquad +b_1a_{12}b_2(\tilde{\mathbf{u}}_1
 \!\cdot\!\tilde{\mathbf{u}}_2)
 \tilde{\mathbf{u}}_1\tilde{\mathbf{u}}_2
 +b_1b_{12}a_2(\tilde{\mathbf{u}}_1
 \!\cdot\!\tilde{\mathbf{u}}_{12})
 \tilde{\mathbf{u}}_1\tilde{\mathbf{u}}_{12}
 \nonumber\\
&\qquad -b_1b_{12}b_2(\tilde{\mathbf{u}}_1
 \!\cdot\!\tilde{\mathbf{u}}_{12})
 (\tilde{\mathbf{u}}_{12}
 \!\cdot\!\tilde{\mathbf{u}}_2)
 \tilde{\mathbf{u}}_1\tilde{\mathbf{u}}_2\bigr]
\end{align}
[$\mathbf{u}_1$ $\!=$ $\!\mathbf{r}$ $\!-$ $\!\mathbf{s}_1$,
$\mathbf{u}_{12}$ $\!=$ $\!\mathbf{s}_1$ $\!-$ $\!\mathbf{s}_2$,
$\mathbf{u}_2$ $\!=$ $\!\mathbf{s}_2$ $\!-$ $\!\mathbf{r}'$;
$q_i$ $\!=$ $\!ku_i$, $a_i$ $\!=$ $\!a(q_i)$, $b_i$ $\!=$ $\!b(q_i)$
for $i$ $\!\in$ $\!\{1,2,12\}$].

\section{Real-cavity model}
\label{subsec_rc}

\begin{figure}[!t!]
\noindent
\begin{center}
\includegraphics[width=0.7\linewidth]{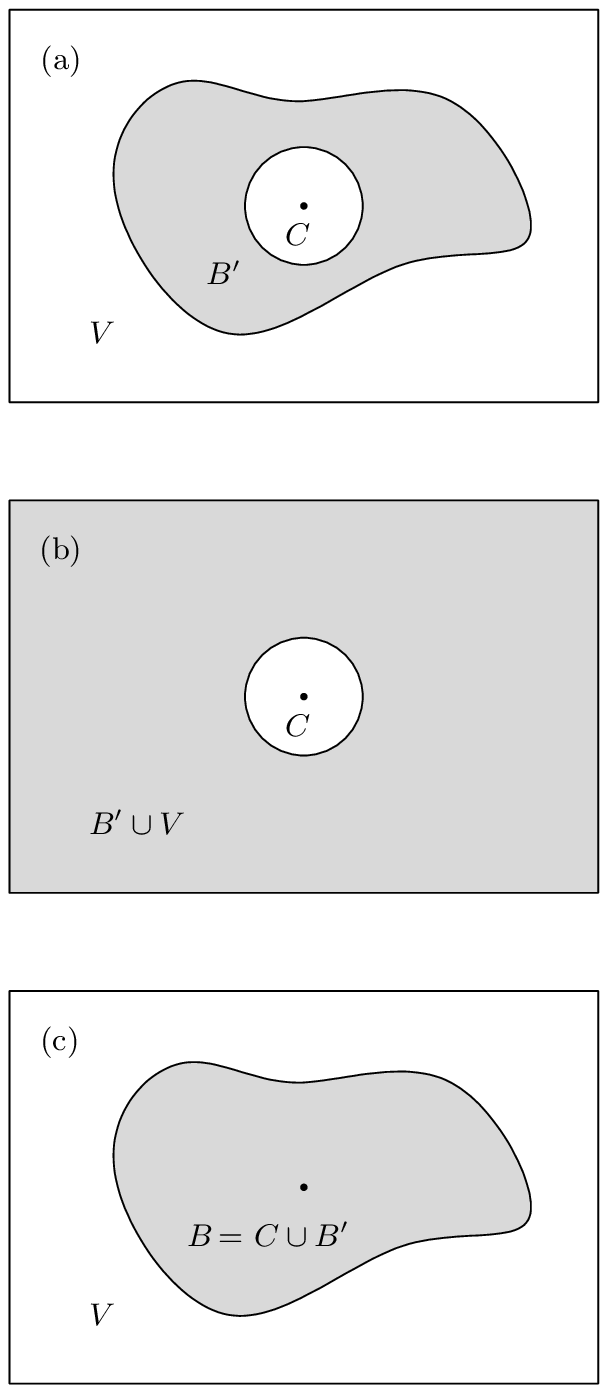}
\end{center}
\caption{
Schematic illustration of the decomposition of (a) $\bm{G}^{(1)}$ into
(b) $\bm{G}^{(1)}_\mathrm{C}$ and (c) $\bm{G}_\mathrm{B}^{(1)}$; the
dot indicates the position of the guest atom.
}
\label{schematic}
\end{figure}

Consider an excited two-level atom embedded in an arbitrary dispersing
and absorbing dielectric body characterized by
$\varepsilon(\mathbf{r},\omega)$. In order to find the
spontaneous-decay rate including local-field corrections, we employ
the real-cavity model, that is to say, we assume that the atom is
located at the center of an empty-space region of the form of a
spherical cavity of radius $R_\mathrm{C}$, as sketched in
Fig.~\ref{schematic}(a), where we have denoted the cavity volume
by $C$, the volume of the host body without the cavity by $B'$ (the
overall volume of the host body being $B$ $\!=$ $\!C\cup B'$), and all
the remaining space by $V$. Hence, the permittivity of the system
changes to
\begin{equation}
\label{e15.1}
\varepsilon_\mathrm{loc}(\mathbf{r},\omega)
=\begin{cases}
1&\mbox{if }\mathbf{r}\in C,\\
\varepsilon(\mathbf{r},\omega)&\mbox{if }\mathbf{r}\not\in C.
\end{cases}
\end{equation}
The cavity radius $R_\mathrm{C}$ is a model parameter representing an
average distance from the atom to the nearest neighboring atoms
constituting the host body; it has to be determined from other
(preferably microscopic) calculations or experiments. Note that the
real-cavity model is applicable provided that the unperturbed host
body is homogeneous (and isotropic) in the region where the guest atom
is implanted,
\begin{equation}
\label{e15.2}
\varepsilon(\mathbf{r},\omega)
=\varepsilon(\mathbf{r}_\mathrm{A},\omega)
\quad\mbox{for }\mathbf{r}\in C.
\end{equation}

\subsection{Linear approximation}
\label{linear}

Restricting our attention to the first-order term in the Born
expansion (\ref{e14.6}), the (scattering) Green tensor corresponding
to $\varepsilon_\mathrm{loc}(\mathbf{r},\omega)$ can be calculated
from Eq.~(\ref{e14.1}), where after some manipulations one obtains
\begin{align}
\label{e15.3}
&\bm{G}^{(1)}
 (\mathbf{r}_\mathrm{A},\mathbf{r}_\mathrm{A},\omega_\mathrm{A})
=\frac{k_\mathrm{A}^4}{16\pi^2}
 \int_{B'}
 \mathrm{d}^3s\,
 \chi(\mathbf{s},\omega_\mathrm{A})e^{(2i-\eta)q}\nonumber\\
&\hspace{20ex}\times[a^2\bm{I}+(b^2-2ab)\tilde{\mathbf{u}}
 \tilde{\mathbf{u}}]\nonumber\\
&\qquad =\bm{G}_\mathrm{C}^{(1)}
 (\mathbf{r}_\mathrm{A},\mathbf{r}_\mathrm{A},\omega_\mathrm{A})
 +\bm{G}_\mathrm{B}^{(1)}
 (\mathbf{r}_\mathrm{A},\mathbf{r}_\mathrm{A},\omega_\mathrm{A})
\end{align}
[$\mathbf{u}$ $\!=$ $\!\mathbf{r}_\mathrm{A}$ $\!-$ $\!\mathbf{s}$,
$q$ $\!=$ $\!k_\mathrm{A}u$], where, again to linear order in the
susceptibility,
\begin{multline}
\label{e15.4}
\bm{G}_\mathrm{C}^{(1)}
 (\mathbf{r}_\mathrm{A},\mathbf{r}_\mathrm{A},\omega_\mathrm{A})
 =\frac{k_\mathrm{A}^4\chi(\omega_\mathrm{A})}{16\pi^2}\\
 \times
 \int_{B'\cup V}\mathrm{d}^3s\,
  e^{(2i-\eta)q}\bigl[a^2\bm{I}+(b^2-2ab)
 \tilde{\mathbf{u}}\tilde{\mathbf{u}}\bigr]
\end{multline}
is the scattering Green tensor of a spherical cavity embedded in a
bulk medium of susceptibility $\chi(\omega_\mathrm{A})$ $\!\equiv$
$\!\chi(\mathbf{r}_\mathrm{A},\omega_\mathrm{A})$
[cf.~Fig.~\ref{schematic}(b)], and
\begin{align}
\label{e15.5}
&\bm{G}_\mathrm{B}^{(1)}
 (\mathbf{r}_\mathrm{A},\mathbf{r}_\mathrm{A},\omega_\mathrm{A})
 =\frac{k_\mathrm{A}^4}{16\pi^2}
 \int_{B}
 \mathrm{d}^3s\,
 \chi(\mathbf{s},\omega_\mathrm{A})e^{(2i-\eta)q}
 \nonumber\\
&\hspace{30ex}\times\bigl[a^2\bm{I}+(b^2-2ab)\tilde{\mathbf{u}}
 \tilde{\mathbf{u}}\bigr]
 \nonumber\\
&-\frac{k_\mathrm{A}^4\chi(\omega_\mathrm{A})}{16\pi^2}
 \int_{B\cup V}
 \mathrm{d}^3s\,
 e^{(2i-\eta)q}\bigl[a^2\bm{I}+(b^2-2ab)
 \tilde{\mathbf{u}}\tilde{\mathbf{u}}\bigr]
\end{align}
is the scattering part of the Green tensor of the host body without
the cavity [cf.~Fig.~\ref{schematic}(c)]. The result can be verified
by applying Eq.~(\ref{e14.2}) [together with Eqs.~(\ref{e8}) and
(\ref{e14.1})] to the two systems mentioned and using
Eq.~(\ref{e15.2}).

Equations (\ref{e15.3})--(\ref{e15.5}) show that $\bm{G}^{(1)}$ in the
rate formula (\ref{e14.4}) can be written as the sum of two terms,
where the first term, $\bm{G}^{(1)}_\mathrm{C}$, only depends on the
cavity radius and the local permittivity of the host body at the
position of the atom, whereas the second term,
$\bm{G}_\mathrm{B}^{(1)}$, is determined by the properties of the host
body in the absence of the atom. Hence, the term in the decay rate
which is proportional to $\mathrm{Im}\,\bm{G}^{(1)}_\mathrm{C}$ can be
regarded as being the local-field correction to the uncorrected term
proportional to $\mathrm{Im}\,\bm{G}^{(1)}_\mathrm{B}$. The fact that
the local-field correction additively enters the rate formula is due
to the linear expansion in $\chi$. Inspection of the second-order term
in the Born expansion, Eq.~(\ref{e16}), indicates that in general,
terms depending on both the cavity and the (unperturbed) host body
will appear which may lead to a breakdown of the additivity. However,
as shown in Sec.~\ref{subsec_con}, there are situations where a
generalization beyond the linear approximation is possible.

Equations (\ref{e15.4}) and (\ref{e15.5}) can be further evaluated
by introducing a spherical coordinate system whose origin coincides
with the position of the atom,
\begin{equation}
\label{e15b}
\int\mathrm{d}^3s \ \rightarrow\
 \int_0^{2\pi}\mathrm{d}\phi\int_0^{\pi}\sin\theta\,\mathrm{d}\theta
 \int_{R_\mathrm{i}(\phi,\theta)}^{R_\mathrm{o}(\phi,\theta)}
 s^2\,\mathrm{d}s,
\end{equation}
where $R_\mathrm{i}(\phi,\theta)$ and $R_\mathrm{o}(\phi,\theta)$,
respectively, refer to the inner and outer boundary areas of the
integration volumes sketched in Figs.~\ref{schematic}(b) and
\ref{schematic}(c). Performing the radial integral in
Eq.~(\ref{e15.4}), we find, on recalling Eq.~(\ref{e14}),
\begin{equation}
\label{e19b1}
 \bm{G}^{(1)}_\mathrm{C}
 (\mathbf{r}_\mathrm{A},\mathbf{r}_\mathrm{A},\omega_\mathrm{A})
 =\lim_{\eta\to 0_+}\lim_{q\to\infty}\bm{F}_\eta(q)
 -\bm{F}_0(q_\mathrm{C})=-\bm{F}_0(q_\mathrm{C})
\end{equation}
($q_\mathrm{C}$ $\!=$ $\!k_\mathrm{A}R_\mathrm{C}$) with
\begin{align}
\label{e19}
&\bm{F}_\eta[q(\phi,\theta)]
 =-\frac{k_\mathrm{A}\chi(\omega_\mathrm{A})}{16\pi^2}
 \int_0^{2\pi}\mathrm{d}\phi\int_0^\pi\mathrm{d}\theta\sin\theta
 \biggl\{e^{(2i-\eta)q}
\nonumber\\
&\times\biggl[\biggl(\frac{1}{3q^3}-\frac{2i}{3q^2}
 -\frac{5}{3q}+\frac{i}{2}\biggr)\bm{I}
 +\biggl(\frac{1}{q^3}-\frac{2i}{q^2}+\frac{3}{q}
 -\frac{i}{2}\biggr)
 \tilde{\mathbf{s}}\tilde{\mathbf{s}}\biggr]
 \nonumber\\
&\hspace{9ex} +4i\,\mathrm{Ei}[(2i\!-\!\eta)q]\biggl(\frac{1}{3}\bm{I}
 -\tilde{\mathbf{s}}\tilde{\mathbf{s}}\biggr)\biggr\}
\end{align}
[$q(\phi,\theta)$ $\!=$ $\!k_\mathrm{A}R(\phi,\theta)$;
$\tilde{\mathbf{s}}$ $\!=$
$\!(\cos\phi\sin\theta,\sin\phi\sin\theta,\cos\theta)$;
$\mathrm{Ei}(x)$, exponential integral]. Using the fact that
$q_\mathrm{C}$ is independent of $\theta$ and $\phi$ as well as the
relation
\begin{equation}
\label{e19b}
\int_0^{2\pi}\mathrm{d}\phi\int_0^\pi\mathrm{d}\theta\,
 \sin\theta\,\tilde{\mathbf{s}}\tilde{\mathbf{s}}
 =\frac{4\pi}{3}\bm{I},
\end{equation}
from Eq.~(\ref{e19}) we obtain
\begin{align}
\label{e19.1}
\bm{F}_0(q_\mathrm{C})
 =-\frac{k_\mathrm{A}\chi(\omega_\mathrm{A})}{12\pi}
 \biggl(\frac{2}{q_\mathrm{C}^3}
 -\frac{4i}{q_\mathrm{C}^2}
 -\frac{2}{q_\mathrm{C}}+i\biggr)e^{2iq_\mathrm{C}}\bm{I}.
\end{align}
In particular, when the radius of the cavity is much smaller
than the atomic transition wavelength,
\begin{equation}
\label{e17}
 q_\mathrm{C}=k_\mathrm{A}R_\mathrm{C}\ll 1,
\end{equation}
then Eq.~(\ref{e19.1}) reduces to
\begin{equation}
\label{e19.2}
\bm{F}_0(q_\mathrm{C})
 =-\frac{k_\mathrm{A}\chi(\omega_\mathrm{A})}{6\pi}
 \biggl[\frac{1}{q_\mathrm{C}^3}+\frac{1}{q_\mathrm{C}}
 +\frac{7i}{6}+O(q_\mathrm{C})\biggr]\bm{I}.
\end{equation}
Substitution of this result together with Eq.~(\ref{e19b1}) into
Eq.~(\ref{e15.3}) reveals that to linear order in $\chi$,
\begin{multline}
\label{e18}
\hspace{-1.5ex}\bm{G}^{(1)}
 (\mathbf{r}_\mathrm{A},\mathbf{r}_\mathrm{A},\omega_\mathrm{A})
 =\frac{k_\mathrm{A}\chi(\omega_\mathrm{A})}{6\pi}
 \biggl[\frac{1}{(k_{\!\mathrm{A}}\!R_\mathrm{C})^3}
 +\frac{1}{k_{\!\mathrm{A}}\!R_\mathrm{C}}+\frac{7i}{6}\biggr]\bm{I}
 \\
+\bm{G}_\mathrm{B}^{(1)}
 (\mathbf{r}_\mathrm{A},\mathbf{r}_\mathrm{A},\omega_\mathrm{A})
 +O(k_\mathrm{A}R_\mathrm{C}).
\end{multline}

For a homogeneous host body which is star-shaped w.r.t. the position
of the atom (i.e., every point on the outer boundary area can be
connected to the atomic position by a straight line that lies entirely
within the body), $\bm{G}_\mathrm{B}^{(1)}$ as given by Eq.~(\ref{e15.5})
can be evaluated in a similar manner. Using again spherical coordinates and
recalling Eq.~(\ref{e14}), we may evaluate the radial integrals in
Eq.~(\ref{e15.5}) to obtain
\begin{align}
\label{e19b2}
&\bm{G}_\mathrm{B}^{(1)}
 (\mathbf{r}_\mathrm{A},\mathbf{r}_\mathrm{A},\omega_\mathrm{A})
 =\lim_{\eta\to 0_+}\bm{F}_\eta[q_\mathrm{o}(\phi,\theta)]
 -\bm{F}_0(0)\nonumber\\
&\quad -\Bigl[\lim_{\eta\to 0_+}\lim_{q\to\infty}
 \bm{F}_\eta(q)-\bm{F}_0(0)\Bigr]
 =\lim_{\eta\to 0_+}\bm{F}_\eta[q_\mathrm{o}(\phi,\theta)]
\end{align}
[$q_\mathrm{o}$ $\!=$ $\!q_\mathrm{o}(\phi,\theta)$ $\!=$
$\!k_\mathrm{A}R_\mathrm{o}(\phi,\theta)$], where henceforth
$R_\mathrm{o}(\phi,\theta)$ denotes the outer boundaries of the host
body.

Substituting Eq.~(\ref{e18}) into Eq.~(\ref{e14.4}), we find that to
linear order in $\chi$ the local-field corrected spontaneous-decay
rate of an atom within a body can be given as follows:
\begin{equation}
\label{e24}
\Gamma
=\Gamma_0+\Gamma_\mathrm{C}+
\Gamma_\mathrm{B}^{(1)},
\end{equation}
where $\Gamma_0$ is the free-space decay rate as given in
Eq.~(\ref{e14.5}),
\begin{align}
\label{e24.1}
\frac{\Gamma_\mathrm{C}}{\Gamma_0}
 =\frac{\mathrm{Im}\,\chi(\omega_\mathrm{A})}
 {(k_\mathrm{A}R_\mathrm{C})^3}
 +\frac{\mathrm{Im}\,\chi(\omega_\mathrm{A})}
 {k_\mathrm{A}R_\mathrm{C}}
 +\frac{7\,\mathrm{Re}\,\chi(\omega_\mathrm{A})}{6},
\end{align}
and
\begin{equation}
\label{e24.2}
\Gamma^{(1)}_\mathrm{B}
 =\frac{2k_\mathrm{A}^2}{\hbar\varepsilon_0}\,
 \mathbf{d}_\mathrm{A}\!\cdot\!\mathrm{Im}\,\bm{G}_\mathrm{B}^{(1)}
 (\mathbf{r}_\mathrm{A},\mathbf{r}_\mathrm{A},\omega_\mathrm{A})
 \!\cdot\!\mathbf{d}_\mathrm{A}.
\end{equation}
Equations (\ref{e24})--(\ref{e24.2}) are valid for an atom embedded in
a weakly polarizable, dispersing and absorbing body of arbitrary size
and shape. Note that only $\Gamma_\mathrm{C}$ takes into account the
effect of local-field correction, while $\Gamma_\mathrm{B}^{(1)}$,
which is simply determined by the scattering part of the Green tensor
of the host body without the cavity, reflects the uncorrected
influence of the size and shape of the body on the decay rate. In
particular, the first two terms in Eq.~(\ref{e24.1}) obviously result
from irreversible energy transfer from the atom to the surrounding
matter. If these two terms are dominant over the last one, effectively
no radiation is emitted.

We conclude this section by making two remarks concerning the
necessary conditions under which Eq.~(\ref{e24}) provides a good
approximation to the spontaneous-decay rate. (i) The permittivity
often appears in the Green tensor as a common factor
$\sqrt{\varepsilon}\omega R_\mathrm{o}/c$ [cf.~Eq.~(\ref{e46})
in Sec.~\ref{subsec_oc}]. To be on the conservative side, one should
then take $\chi k_\mathrm{A}R_\mathrm{max}$ [with $R_\mathrm{max}$
$\!=$ $\max_{(\theta,\phi)}R_\mathrm{o}(\theta,\phi)$] rather than
$\chi$ as the small parameter. (ii) Equation (\ref{e24.1}) already
contains terms of the order of
$\mathrm{Im}\,\chi/(k_\mathrm{A}R_\mathrm{C})^3$. Thus
$\mathrm{Im}\,\chi/(k_\mathrm{A}R_\mathrm{C})^3$, not
$\mathrm{Im}\,\chi$, should be much smaller than unity for the
linear approximation to yield a good estimation of the decay rate. The
cavity radius  $R_\mathrm{C}$ represents the average distance between
the atom and the constituents of the body. For, say,
$k_\mathrm{A}R_\mathrm{C}$ $\!\gtrsim$ $\!0.01$, it is sufficient
to require that \mbox{$\mathrm{Im}\,\chi$ $\!\ll$ $\!10^{-6}$}, so
that \mbox{$\mathrm{Im}\,\chi/(k_\mathrm{A}R_\mathrm{C})^3$
$\!\ll$ $\!1$}.

\subsection{Beyond the linear approximation}
\label{subsec_con}

\begin{figure}[!t!]
\noindent
\includegraphics[width=.6\linewidth]{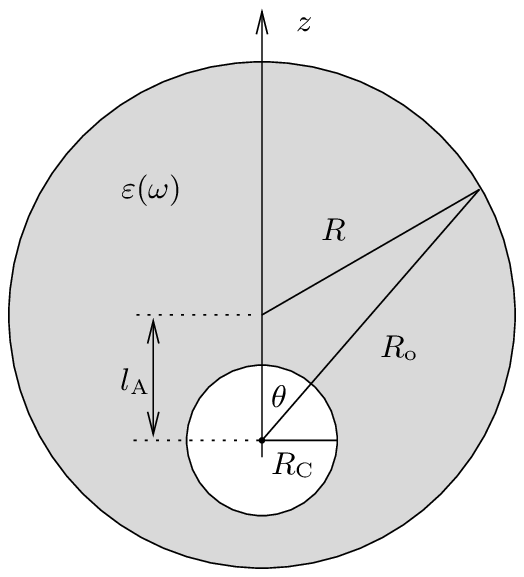}
\caption{Real-cavity model for an atom embedded in a homogeneous
dielectric sphere.}
\label{sphere}
\end{figure}

Toma\v{s} \cite{Tomas01} has found that in the special case of an atom
situated at the center of a homogenous, dielectric sphere, the
real-cavity model (Fig.~\ref{sphere} in the case where
\mbox{$l_\mathrm{A}$ $\!=$ $\!0$}) leads to
the relation
\begin{align}
\label{e21}
&\bm{G}^{(1)}
 (\mathbf{r}_\mathrm{A},\mathbf{r}_\mathrm{A},\omega_\mathrm{A})
 =\frac{k_\mathrm{A}}{6\pi}\biggl\{
 \frac{3(\varepsilon-1)}{2\varepsilon+1}\,
 \frac{1}{(k_\mathrm{A}R_\mathrm{C})^3}
\nonumber\\
&\quad+\frac{9(\varepsilon-1)(4\varepsilon+1)}{5(2\varepsilon+1)^2}\,
 \frac{1}{k_\mathrm{A}R_\mathrm{C}}
 +i\biggl[\frac{9\varepsilon^{5/2}}{(2\varepsilon+1)^2}
 -1\biggr]\biggr\}\bm{I}\nonumber\\
&\quad+\biggl(\frac{3\varepsilon}{2\varepsilon+1}\biggr)^2
 \bm{G}_\mathrm{B}^{(1)}
 (\mathbf{r}_\mathrm{A},\mathbf{r}_\mathrm{A},\omega_\mathrm{A})
 +O(k_\mathrm{A}R_\mathrm{C}),
\end{align}
$\varepsilon$ $\!=$ $\varepsilon(\omega_\mathrm{A})$, where
$\bm{G}_\mathrm{B}^{(1)}
(\mathbf{r},\mathbf{r}',\omega)$ is the scattering Green tensor of the
homogeneous dielectric sphere, whose geometry dependence is---for
$\mathbf{r}$ $\!=$ $\!\mathbf{r}'$ $\!=$ $\!\mathbf{r}_\mathrm{A}$ at
the sphere center---entirely given by its dependence on the sphere
radius $R$. As already mentioned in Sec.~\ref{sec_intro}, he made the
conjecture that this relation might be more generally valid for (i)
bodies of arbitrary sizes and shapes and (ii) arbitrary positions of
the atom, provided that the atom is not on the body surface.

Let us study the validity of this conjecture in more detail. It can
easily be seen that to linear order in $\chi$, Eq.~(\ref{e21})
obviously reduces to Eq.~(\ref{e18}), so the results of
Sec.~\ref{linear} show that in this order the conjecture is true even
under the more general conditions of inhomogeneous host bodies,
provided that the requirement (\ref{e15.2}) for the applicability of
the real-cavity model is valid. Moreover, we will demonstrate in the
following that Eq.~(\ref{e21}) remains valid beyond the linear order
in $\chi$, provided that the respective host body can be regarded as
being homogeneous in the vicinity of the atom, i.e.,
\begin{equation}
\label{e15.2b}
\varepsilon(\mathbf{r},\omega_\mathrm{A})
 =\varepsilon(\mathbf{r}_\mathrm{A},\omega_\mathrm{A})
 \quad\mbox{for }|\mathbf{r}-\mathbf{r}_\mathrm{A}|\le
 (1+\nu)R_\mathrm{C},
\end{equation}
with $\nu$ being some small positive number.

We begin with the case of a homogeneous body, in which case
satisfaction of the condition (\ref{e15.2b}) simply ensures that the
entire cavity lies inside the body. We first recall that
$\bm{G}^{(1)}
(\mathbf{r}_\mathrm{A},\mathbf{r}_\mathrm{A},\omega_\mathrm{A})$
determines the electric field reaching the point
$\mathbf{r}_\mathrm{A}$, in which it has originated, after being
scattered at the surfaces of inhomogeneity
[cf.~Fig.~\ref{schematic}(a)]. At the cavity surface the electric
field can be separated into two parts, namely, one part that is
(multiply) reflected at this surface and eventually returns to the
point $\mathbf{r}_\mathrm{A}$, and one part that is eventually
transmitted to the exterior of the cavity. The contribution due to the
reflected part is obviously given by
$\bm{G}^{(1)}_\mathrm{C}(\mathbf{r}_\mathrm{A},\mathbf{r}_\mathrm{A},
\omega_\mathrm{A})$ [cf.~Fig.~\ref{schematic}(b)], which reads
\cite{Scheel99b,Tomas01}
\begin{align}
\label{e21c}
&\bm{G}^{(1)}_\mathrm{C}
 (\mathbf{r}_\mathrm{A},\mathbf{r}_\mathrm{A},\omega_\mathrm{A})
 =\frac{k_\mathrm{A}}{6\pi}\biggl\{
 \frac{3(\varepsilon-1)}{2\varepsilon+1}\,
 \frac{1}{(k_\mathrm{A}R_\mathrm{C})^3}
\nonumber\\
&\quad+\frac{9(\varepsilon-1)(4\varepsilon+1)}{5(2\varepsilon+1)^2}\,
 \frac{1}{k_\mathrm{A}R_\mathrm{C}}
 +i\biggl[\frac{9\varepsilon^{5/2}}{(2\varepsilon+1)^2}
 -1\biggr]\biggr\}\bm{I}\nonumber\\
&\quad+O(k_\mathrm{A}R_\mathrm{C}).
\end{align}
In an infinitely extended body [cf.~Fig.~\ref{schematic}(b)], the
transmitted part at a point $\mathbf{r}$ outside the cavity
is determined by the Green tensor \cite{Li94}
\begin{equation}
\label{e21d}
\bm{G}_\mathrm{C}(\mathbf{r},\mathbf{r}_\mathrm{A},\omega_\mathrm{A})
 =\frac{k_\mathrm{A}A}{4\pi}\,(a\bm{I}
 -b\tilde{\mathbf{u}}\tilde{\mathbf{u}})e^{iq}
\end{equation}
[$\mathbf{u}$ $\!=$ $\!\mathbf{r}$ $\!-$ $\!\mathbf{r}_\mathrm{A}$,
$q$ $\!=$ $\!nk_\mathrm{A}u$, $n$ $\!=$ $\sqrt{\varepsilon}$,
and $a$ and $b$ according to Eq.~(\ref{e14})]. The coefficient $A$
(which corresponds to $A_N^{12}$ in Ref.~\cite{Li94}) is given by
\begin{equation}
\label{e21e}
A=n\,\frac{j_1(z_0)[z_0h_1^{(1)}(z_0)]'
 -[z_0j_1(z_0)]'h_1^{(1)}(z_0)}{j_1(z_0)[z_1h_1^{(1)}(z_1)]'
 -\varepsilon[z_0j_1(z_0)]'h_1^{(1)}(z_1)}\,,
\end{equation}
where $z_0$ $\!=$ $\!k_\mathrm{A}R_\mathrm{C}$, $z_1$ $\!=$
$\!nk_\mathrm{A}R_\mathrm{C}$ (the primes denote derivatives w.r.t.
$z_0$ and $z_1$), and
\begin{equation}
\label{e2.1}
j_1(z)=\frac{\sin z}{z^2}-\frac{\cos z}{z}
\end{equation}
and
\begin{equation}
\label{e47}
h_1^{(1)}(z)=-\biggl(\frac{1}{z}+\frac{i}{z^2}\biggr)e^{iz}
\end{equation}
are the spherical Bessel and Hankel functions, respectively.
Inspection of Eq.~(\ref{e21e}) shows that
\begin{equation}
\label{e21f}
A=n\,\frac{3\varepsilon}{2\varepsilon+1}+O(k_\mathrm{A}R_\mathrm{C}).
\end{equation}
Substitution of this result into Eq.~(\ref{e21d}) leads to
\begin{equation}
\label{e21g}
\bm{G}_\mathrm{C}(\mathbf{r},\mathbf{r}_\mathrm{A},\omega_\mathrm{A})
 =\frac{3\varepsilon}{2\varepsilon+1}\,\bm{G}_\mathrm{B}^{(0)}
 (\mathbf{r},\mathbf{r}_\mathrm{A},\omega_\mathrm{A})
 +O(k_\mathrm{A}R_C),
\end{equation}
where $\bm{G}_\mathrm{B}^{(0)}
(\mathbf{r},\mathbf{r}_\mathrm{A},\omega_\mathrm{A})$ is the Green
tensor of the infinite body without the cavity [according to
Eqs.~(\ref{e13}) and (\ref{e14}) with $k \mapsto n k$]. In other
words, the electric field transmitted through the cavity surface to a
point outside the cavity is equal to the field that would be
transmitted to the same point in the absence of the cavity, multiplied
by a global factor. By means of the general symmetry property (see,
e.g., Ref.~\cite{Knoell01})
\begin{equation}
\label{e21h}
\bm{G}(\mathbf{r},\mathbf{r}',\omega)
 =\bm{G}^\mathrm{T}(\mathbf{r}',\mathbf{r},\omega)
\end{equation}
we also have
\begin{equation}
\label{e21i}
\bm{G}_\mathrm{C}(\mathbf{r}_\mathrm{A},\mathbf{r},\omega_\mathrm{A})
 =\frac{3\varepsilon}{2\varepsilon+1}\,
 \bm{G}_\mathrm{B}^{(0)}
 (\mathbf{r}_\mathrm{A},\mathbf{r},\omega_\mathrm{A})
 +O(k_\mathrm{A}R_\mathrm{C}),
\end{equation}
i.e., the electric field transmitted through the cavity surface from
an arbitrary point outside the cavity is also equal to the
corresponding result in the absence of the cavity, multiplied by the
same factor.

For a finite body, the electric field will be (multiply) reflected
from the body's outer surface, eventually giving rise to a field at
$\mathbf{r}_\mathrm{A}$. Without the cavity, processes of this kind
are taken into account by replacing the infinite-body Green tensor
$\bm{G}_\mathrm{B}^{(0)}$ with its finite-body counterpart
$\!\bm{G}_\mathrm{B}$. Combining this observation with
Eqs.~(\ref{e21g}) and (\ref{e21i}), we conclude, on recalling the
linearity of Maxwell's equations, that the electric field which is
transmitted though the cavity surface, scattered at the outer body
surface, and finally retransmitted into the cavity is given by
\begin{equation}
\label{e21j}
\biggl(\frac{3\varepsilon}{2\varepsilon+1}\biggr)^2\,
 \bm{G}_\mathrm{B}^{(1)}
 (\mathbf{r}_\mathrm{A},\mathbf{r}_\mathrm{A},\omega_A)
 +O(k_\mathrm{A}R_\mathrm{C}),
\end{equation}
so combining Eqs.~(\ref{e21c}) and (\ref{e21j}), we arrive at
Eq.~(\ref{e21}).

In this derivation, we have disregarded processes involving scattering
of the field at the cavity surface from the outside. Processes of this
kind can indeed be neglected, because their contributions are of
orders higher than $O(k_\mathrm{A}R_\mathrm{C})$. To see this, note
that for a cavity in bulk material [Fig.~\ref{schematic}(b)] the field
at a point $\mathbf{r}$ outside the cavity originating in a point
$\mathbf{r}'$ outside the cavity via scattering at the cavity surface,
is characterized by \cite{Li94}
\begin{multline}
\label{e21k}
\bm{G}^{(1)}_\mathrm{C}
 (\mathbf{r},\mathbf{r}',\omega_\mathrm{A})\\
 =\sum_{m=1}^\infty \Big[B_m^M
 \bm{M}(\mathbf{r},\mathbf{r}',\omega_\mathrm{A})
 +B_m^N
 \bm{N}(\mathbf{r},\mathbf{r}',\omega_\mathrm{A})\Big],
\end{multline}
where
\begin{align}
\label{e21l}
&B_m^M=-\frac{j_m(z_0)[z_1j_m(z_1)]'
 -[z_0j_m(z_0)]'j_m(z_1)}{j_m(z_0)[z_1h_m^{(1)}(z_1)]'
 -[z_0j_m(z_0)]'h_m^{(1)}(z_1)}\,,\\
&B_m^N=-\frac{j_m(z_0)[z_1j_m(z_1)]'
 -\varepsilon[z_0j_m(z_0)]'j_m(z_1)}{j_m(z_0)[z_1h_m^{(1)}(z_1)]'
 -\varepsilon[z_0j_m(z_0)]'h_m^{(1)}(z_1)},
\end{align}
and the tensors $\bm{M}(\mathbf{r},\mathbf{r}',\omega_\mathrm{A})$
and $\bm{N}(\mathbf{r},\mathbf{r}',\omega_\mathrm{A})$ do not depend
on $R_\mathrm{C}$. Using the relations \cite{Abramowitz73}
\begin{align}
\label{e21m}
&j_m(z) = \frac{z^m}{(2m+1)!!}\bigl[1+O(z^2)\bigr],\\
\label{e21n}
&h_m^{(1)}(z) = -\frac{i(2m-1)!!}{z^{m+1}}
 \bigl[1+O(z^2)\bigr],
\end{align}
one can easily show that
\begin{equation}
\label{e21o}
B_m^M=O\bigl[(k_\mathrm{A}R_\mathrm{C})^{2m+3}\bigr],\quad
B_m^N=O\bigl[(k_\mathrm{A}R_\mathrm{C})^{2m+1}\bigr],
\end{equation}
so Eq.~(\ref{e21k}) leads to
\begin{equation}
\label{e21p}
\bm{G}^\mathrm{(1)}_\mathrm{C}
 (\mathbf{r},\mathbf{r}',\omega_\mathrm{A})
 =O\bigl[(k_\mathrm{A}R_\mathrm{C})^3\bigr].
\end{equation}
Processes involving reflections of the field at the cavity surface
from the outside hence start to contribute in third order of
$k_\mathrm{A}R_\mathrm{C}$ and therefore do not need to be included in
Eq.~(\ref{e21}).

So far we have demonstrated the validity of the relation (\ref{e21})
for homogeneous dielectric bodies of arbitrary shapes provided that
the atom is situated at the interior of the body, such that the
condition (\ref{e15.2b}) is satisfied. Note that this condition
practically coincides with the condition (\ref{e15.2}) for the
applicability of the real-cavity model. The arguments given above can
be extended to inhomogeneous bodies. If the condition (\ref{e15.2b})
is satisfied, one can divide such a body into a more or less small
homogeneous part containing the cavity plus an inhomogeneous rest.
Equations~(\ref{e21g}), (\ref{e21i}), and (\ref{e21p}) then still
describe the propagation of the electric field inside the homogeneous
part of the body, and the effect of the inhomogeneous part can be
taken into account by the scattering at the (fictitious) surface
dividing the two parts. Consequently, we are again left with
Eq.~(\ref{e21}).

Substituting Eq.~(\ref{e21}) into Eq.~(\ref{e14.4}) and recalling
Eq.~(\ref{e14.5}), we can again cast the local-field corrected
spontaneous-decay rate in the form of of Eq.~(\ref{e24}), where now
\begin{align}
\label{e22.2}
\frac{\Gamma_\mathrm{C}}{\Gamma_0}
 =&\;\mathrm{Im}\biggl\{
 \frac{3(\varepsilon-1)}{2\varepsilon+1}\,
 \frac{1}{(k_\mathrm{A}R_\mathrm{C})^3}
 +\frac{9(\varepsilon-1)(4\varepsilon+1)}{5(2\varepsilon+1)^2}\,
 \frac{1}{k_\mathrm{A}R_\mathrm{C}}\nonumber\\
 &\;+i\biggl[\frac{9\varepsilon^{5/2}}{(2\varepsilon+1)^2}
 -1\biggr]\biggr\},
\end{align}
and
\begin{equation}
\label{e22.2b}
\Gamma_\mathrm{B}^{(1)}
 =\frac{2k_\mathrm{A}^2}{\hbar\varepsilon_0}\,
 \mathbf{d}_\mathrm{A}\!\cdot\!\mathrm{Im}\biggl
[
 \biggl(\frac{3\varepsilon}{2\varepsilon+1}\biggr)^2
 \bm{G}_\mathrm{B}^{(1)}
 (\mathbf{r}_\mathrm{A},\mathbf{r}_\mathrm{A},\omega_\mathrm{A})
 \biggr
]
\!\cdot\mathbf{d}_\mathrm{A}
\end{equation}
[$\varepsilon$ $\!=$ $\varepsilon(\mathbf{r}_\mathrm{A},
\omega_\mathrm{A})$]. Recall that $\bm{G}_\mathrm{B}^{(1)}$ is the
(unperturbed) scattering part of the Green tensor of the host body
without the cavity. Needless to say that to linear order in~$\chi$,
Eqs.~(\ref{e22.2}) and (\ref{e22.2b}) reduce to Eqs.~(\ref{e24.1}) and
(\ref{e24.2}), respectively.

In particular in the case of weakly absorbing matter, it may be
sufficient to retain in Eqs.~(\ref{e22.2}) and (\ref{e22.2b}) only
terms to linear order in $\mathrm{Im}\,\varepsilon$. From
Eqs.~(\ref{e24}), (\ref{e22.2}), and (\ref{e22.2b}) it then follows
that when
\begin{equation}
\label{Eq.60}
\left|\frac{\mathrm{Im}\,\varepsilon\,\mathbf{d}_\mathrm{A}
\cdot\mathrm{Re}\,\bm{G}_\mathrm{B}^{(1)}
(\mathbf{r}_\mathrm{A},\mathbf{r}_\mathrm{A},\omega_\mathrm{A})
\cdot\mathbf{d}_\mathrm{A}}
{\mathrm{Re}\,\varepsilon\,\mathbf{d}_\mathrm{A}
\cdot\mathrm{Im}\,\bm{G}_\mathrm{B}
(\mathbf{r}_\mathrm{A},\mathbf{r}_\mathrm{A},\omega_\mathrm{A})
\cdot\mathbf{d}_\mathrm{A}}\right| \ll 1
\end{equation}
then $\Gamma$ can be given in the form
\begin{equation}
\label{Eq.61}
\Gamma
 =\left(\frac{3\,\mathrm{Re}\,\varepsilon}
 {2\,\mathrm{Re}\,\varepsilon+1}\right)^2\Gamma_\mathrm{B}
 +\Delta\Gamma,
\end{equation}
where
\begin{align}
\label{Eq.62}
\frac{\Delta\Gamma}{\Gamma_0}
 =&\,\frac{9}
 {(2\,\mathrm{Re}\,\varepsilon+1)^2}\,
 \frac{\mathrm{Im}\,\varepsilon}{(k_\mathrm{A}R_\mathrm{C})^3}
 \nonumber\\
&+\frac{9(14\mathrm{Re}\,\varepsilon+1)}
 {5(2\,\mathrm{Re}\,\varepsilon+1)^3}\,
 \frac{\mathrm{Im}\,\varepsilon}{k_\mathrm{A}R_\mathrm{C}}\,,
\end{align}
and $\Gamma_\mathrm{B}$ is the uncorrected decay rate as given by
Eq.~(\ref{Eq.10}) with $\bm{G}^{(1)}
(\mathbf{r}_\mathrm{A},\mathbf{r}_\mathrm{A},\omega_\mathrm{A})$
$\!\equiv$ $\bm{G}_\mathrm{B}^{(1)}
(\mathbf{r}_\mathrm{A},\mathbf{r}_\mathrm{A},\omega_\mathrm{A})$.
Rahmani and Bryant \cite{Rahmani02} concluded from a numerical
computation of the Green tensor of a dielectric sphere that contains a
(small) empty sphere at an arbitrary position inside the material that
the local-field corrected spontaneous decay rate has the form
$\Gamma$ $\!=$ $f^2\Gamma_\mathrm{B}$ $\!+$ $\delta\Gamma$,
where the shift term $\delta\Gamma$ is due to absorption. Comparing
with Eq.~(\ref{Eq.61}) [together with Eq.~(\ref{Eq.62})], we see that
this is indeed the case when the effect of absorption is sufficiently
weak and in particular, the inequality (\ref{Eq.60}) holds. However,
from Eqs.~(\ref{e24}), (\ref{e22.2}), and (\ref{e22.2b}) it is clearly
seen that in general, $\Gamma$ cannot be given in the form assumed in
Ref.~\cite{Rahmani02}. Note that already the analytical solution for
the special case considered in Ref.~\cite{Tomas01} implies that this
form cannot be true in general.

\section{Examples}
\label{sec_ex}

\subsection{Atom in a bulk medium}
\label{subsec_bulk}

In the case of bulk material, we have $\bm{G}_\mathrm{B}^{(1)}$ $\!=$
$\!0$, so Eq.~(\ref{e24}) simplifies to
\begin{equation}
\label{e23s1}
\Gamma= \Gamma_0 + \Gamma_\mathrm{C},
\end{equation}
where $\Gamma_\mathrm{C}/\Gamma_0$ is given by Eq.~(\ref{e22.2})
\cite{Scheel99b,Tomas01}, which to linear order in $\chi$ reduces to
Eq.~(\ref{e24.1}). In particular, when material absorption can be
neglected, $\varepsilon$ $\!\simeq$ $\!\mathrm{Re}\,\varepsilon$, we
simply have~\cite{Glauber91}
\begin{equation}
\label{e0}
\Gamma=\biggl(\frac{3\varepsilon}{2\varepsilon+1}\biggr)^2
  \sqrt{\varepsilon}\,\Gamma_0
\end{equation}
[which follows directly from Eq.~(\ref{Eq.61}) with
$\Gamma_\mathrm{B}$ $\!=$ $\!\sqrt{\varepsilon}\Gamma_0$,
cf.~Eq.~(\ref{Eq.10})]. Note
that in this case the virtual-cavity model leads to
\begin{equation}
\label{e0a}
\Gamma
 =\biggl(\frac{\varepsilon+2}{3}\biggr)^2
 \sqrt{\varepsilon}\,\Gamma_0
\end{equation}
(see, e.g., Ref.~\cite{Scheel99a}). It is not difficult to see that
to linear order in $\chi$ both Eq.~(\ref{e0}) and Eq.~(\ref{e0a})
lead to
\begin{equation}
\label{e1}
\Gamma
 =\biggl[1+\frac{7\chi(\omega_\mathrm{A})}{6}
 \biggr]\Gamma_0,
\end{equation}
i.e., the first-order theory for the spontaneous decay rate does not
distinguish between the virtual- and the real-cavity model
\cite{Berman04}, provided that absorption can be disregarded.

\subsection{Atom in a sphere}
\label{subsec_SP}

Let us consider an atom in a homogeneous dielectric sphere as sketched
in Fig.~\ref{sphere} and first calculate the spontaneous-decay rate to
linear order in $\chi$. In particular, we need to calculate
$\Gamma_\mathrm{B}^{(1)}$ as given by Eq.~(\ref{e24.2}) with
$\bm{G}_\mathrm{B}^{(1)}$ from Eq.~(\ref{e19b2}) together with
Eq.~(\ref{e19}). Noting that
\begin{equation}
\label{e25}
q_\mathrm{o}(\phi,\theta)=k_\mathrm{A}\Bigl[l_\mathrm{A}\cos\theta
 +\sqrt{R^2-l_\mathrm{A}^2(1-\cos^2\theta)}\Bigl]
\end{equation}
(cf.~Fig.~\ref{sphere}) is independent of $\phi$,
performing the $\phi$-integral, and using the fact that
$\mathbf{d}_\mathrm{A}\cdot\tilde{\mathbf{s}}$ $\!=$
$\!d_\mathrm{A}\cos\theta$ for a radially ($\perp$) oriented dipole
and $\mathbf{d}_\mathrm{A}\cdot\tilde{\mathbf{s}}$
$\!=$ $\!d_\mathrm{A}\sin\theta\cos\phi$ for a tangentially
($\parallel$) oriented dipole, we find
\begin{align}
\label{e26}
\Gamma_\mathrm{B}^{(1)\perp(\parallel)}
= -\frac{3\Gamma_0}{4}\,\lim_{\eta\to 0}\mathrm{Im}
 \biggl[\chi(\omega_\mathrm{A})\int_{-1}^1\mathrm{d}x \,
 f_\eta^{\perp(\parallel)}(q_\mathrm{o},x^2)
 \biggr],
\end{align}
where
\begin{align}
\label{e27}
f_\eta^\perp(q_\mathrm{o},z)
 =&\;\biggl[
 \frac{1}{3q_\mathrm{o}^3}-\frac{2i}{3q_\mathrm{o}^2}
 -\frac{5}{3q_\mathrm{o}}+\frac{i}{2}
 \nonumber\\
&\quad +\biggl(\frac{1}{q_\mathrm{o}^3}-\frac{2i}{q_\mathrm{o}^2}
 +\frac{3}{q_\mathrm{o}}-\frac{i}{2}\biggr)z\biggr]
 e^{(2i-\eta)q_\mathrm{o}}\nonumber\\
&+4i\biggl(\frac{1}{3}-z\biggr)
 \mathrm{Ei}[(2i-\eta)q_\mathrm{o}],\\[1ex]
\label{e28}
f_\eta^\parallel(q_\mathrm{o},z)
 =&\;
f_\eta^\perp[q_\mathrm{o},(1-z)/2],
\end{align}
and $q_\mathrm{o}$ according to Eq.~(\ref{e25}), with $\cos\theta$
being replaced by~$x$.

In order to exactly calculate $\Gamma_\mathrm{B}^{(1)}$
[Eq.~(\ref{e22.2b})], we make use of the exact Green tensor for a
dielectric sphere as given in Ref.~\cite{Li94}, leading to
\begin{multline}
\label{e34}
\Gamma_\mathrm{B}^{(1)\perp}
 =\frac{3\Gamma_0}{2}\,\mathrm{Im}\biggl\{
 \frac{9i\varepsilon^{5/2}}{(2\varepsilon+1)^2}
 \sum_{m=1}^\infty(2m+1)\\
\times m(m+1)C^N_m\biggl[\frac{j_m(nk_\mathrm{A}l_\mathrm{A})}
 {nk_\mathrm{A}l_\mathrm{A}}\biggr]^2
\biggr\},
\end{multline}
\begin{multline}
\label{e35}
\Gamma_\mathrm{B}^{(1)\|}
 =\frac{3\Gamma_0}{4}\,\mathrm{Im}\biggl\{
 \frac{9i\varepsilon^{5/2}}{(2\varepsilon+1)^2}
 \sum_{m=1}^\infty(2m+1)\\
\times \biggl[C^M_m j_m^2(nk_\mathrm{A}l_\mathrm{A})+C^N_m
 \biggl(\frac{\mathrm{d}[l_\mathrm{A}j_m(nk_\mathrm{A}l_\mathrm{A})]}
 {(nk_\mathrm{A}l_\mathrm{A})\mathrm{d}l_\mathrm{A}}\biggr)^2
 \biggr]\biggr\},
\end{multline}
where
\begin{align}
\label{e36}
&C^N_m=-\frac{\varepsilon h_m^{(1)}(z_1)[z_0h_m^{(1)}(z_0)]'
 -[z_1h_m^{(1)}(z_1)]'h_m^{(1)}(z_0)}
 {\varepsilon j_m^{(1)}(z_1)[z_0h_m^{(1)}(z_0)]'
 -[z_1j_m^{(1)}(z_1)]'h_m^{(1)}(z_0)}\,,\\
\label{e37}
&C^M_m=-\frac{h_m^{(1)}(z_1)[z_0h_m^{(1)}(z_0)]'
 -[z_1h_m^{(1)}(z_1)]'h_m^{(1)}(z_0)}
 {j_m^{(1)}(z_1)[z_0h_m^{(1)}(z_0)]'
 -[z_1j_m^{(1)}(z_1)]'h_m^{(1)}(z_0)}\,,
\end{align}
with $z_0$ $\!\equiv$ $\!k_\mathrm{A}R$, $z_1$ $\!\equiv$
$\!nk_\mathrm{A}R$.

\subsubsection{Atom on center}
\label{subsec_oc}

To compare the exact decay rate [Eq.~(\ref{e24}) together with
Eqs.~(\ref{e14.5}), and (\ref{e22.2}) as well as Eqs.~(\ref{e34})
and/or (\ref{e35})] with the decay rate obtained in the linear Born
approximation [Eq.~(\ref{e24}) together with Eqs.~(\ref{e14.5}),
(\ref{e24.1}), and (\ref{e26})], let us consider, for simplicity, the
case where the atom is positioned at the center of the sphere. Setting
$l_\mathrm{A}$ $\!=$ $\!0$ and hence, \mbox{$q_\mathrm{o}$ $\!=$
$\!k_\mathrm{A}R$} in Eqs.~(\ref{e26})--(\ref{e28}), and carrying out
the $x$ integral in Eq.~(\ref{e26}), we derive
\begin{equation}
\label{e43.1}
\Gamma_\mathrm{B}^{(1)\perp}=\Gamma_\mathrm{B}^{(1)\parallel}
=\Gamma_\mathrm{B}^{(1)}
\end{equation}
with
\begin{multline}
\label{e44}
\Gamma_\mathrm{B}^{(1)}
 =-\Gamma_0\lim_{\eta\to 0}\mathrm{Im}\biggl\{\chi(\omega_\mathrm{A})
 \biggl[\frac{1}{(k_\mathrm{A}R)^3}
 -\frac{2i}{(k_\mathrm{A}R)^2}\\
-\frac{1}{k_\mathrm{A}R}
 +\frac{i}{2}\biggr]e^{(2i-\eta)k_\mathrm{A}R}\biggr\},
\end{multline}
giving
\begin{multline}
\label{e44.1}
\Gamma_\mathrm{B}^{(1)}
 =-\Gamma_0\mathrm{Im}\biggl\{\chi(\omega_\mathrm{A})
 \biggl[\frac{1}{(k_\mathrm{A}R)^3}
 -\frac{2i}{(k_\mathrm{A}R)^2}\\
 -\frac{1}{k_\mathrm{A}R}
 +\frac{i}{2}\biggr]e^{2ik_\mathrm{A}R}\biggr\}
\end{multline}
for any finite radius $R$, and approaching zero in the limit $R$
$\!\to$ $\!\infty$ (which has to be taken before performing the limit
$\eta$ $\!\to$ $\!0$). When material absorption is small enough such
that
$\mathrm{Im}\,\chi(\omega_\mathrm{A})/(k_\mathrm{A}R_\mathrm{C})^3$
$\!\ll$ $\!7\mathrm{Re}\,\chi(\omega_\mathrm{A})/6$,
and the radius of the sphere is large $k_\mathrm{A}R$ $\!\gg$ $\!1$,
then Eq.~(\ref{e24}) together with Eqs.~(\ref{e24.1}) and
(\ref{e44.1}) reduces to [$\chi(\omega_\mathrm{A})$ $\!\simeq$
$\!\mathrm{Re}\,\chi(\omega_\mathrm{A})$]
\begin{equation}
\label{e2}
\Gamma=\biggl[1+\frac{7\chi(\omega_\mathrm{A})}{6}
 \biggr]\Gamma_0
 -\frac{\chi(\omega_\mathrm{A})}{2}\cos(2k_\mathrm{A}R)\Gamma_0.
\end{equation}
Note that Eq.~(\ref{e2}) differs from Eq.~(\ref{e1}) in the second
term, which reflects the finite size of the sphere.

When the atom is on center, only the terms \mbox{$m$ $\!=$ $\!1$}
in Eqs.~(\ref{e34}) and (\ref{e35}) contribute to the exact
$\Gamma^{(1)}$, cf.~Eqs.~(\ref{e21m}) and (\ref{e21n}), and hence
we find
\begin{equation}
\label{e45}
\Gamma_\mathrm{B}^{(1)}
 =\Gamma_0\mathrm{Im}\biggl[
 \frac{9i\varepsilon^{5/2}}{(2\varepsilon+1)^2}\,
 C^N_1\biggr].
\end{equation}
Combining Eqs.~(\ref{e24}),
(\ref{e22.2}), and (\ref{e45}), we arrive at
\begin{align}
\label{Eq.80}
\Gamma=
&\;\mathrm{Im}\biggl\{
 \frac{3(\varepsilon-1)}{2\varepsilon+1}\,
 \frac{1}{(k_\mathrm{A}R_\mathrm{C})^3}
 +\frac{9(\varepsilon-1)(4\varepsilon+1)}{5(2\varepsilon+1)^2}\,
 \frac{1}{k_\mathrm{A}R_\mathrm{C}}
 \nonumber\\
&\;
 +\frac{9i\varepsilon^{5/2}}{(2\varepsilon+1)^2}\,\bigl(1+C^N_1\bigr)
 \biggr\}\Gamma_0,
\end{align}
which can be shown to agree with Eq.~(\ref{e44.1}) to linear order in
$\chi$. In a more sophisticated linearization of Eq.~(\ref{Eq.80})
with respect to $\chi$, it may be advantageous to leave the
$\chi$-dependence in the exponentials appearing in the coefficient
$C^N_1$ unchanged. In particular, for sufficiently small material
absorption,
$\mathrm{Im}\,\chi(\omega_\mathrm{A})/(k_\mathrm{A}R_\mathrm{C})^3$
$\!\ll$ $\!7\mathrm{Re}\,\chi(\omega_\mathrm{A})/6$, and for a
sufficiently large sphere, $k_\mathrm{A}R$ $\!\gg$ $\!1$, this kind of
linear approximation leads to [$\chi(\omega_\mathrm{A})$ $\!\simeq$
$\!\mathrm{Re}\,\chi(\omega_\mathrm{A})$]
\begin{multline}
\label{e46}
\Gamma
 = \biggl[1+\frac{7\chi(\omega_\mathrm{A})}{6}\biggr]\Gamma_0\\
 -\frac{\chi(\omega_\mathrm{A})}{2}\,
 \cos\bigl[2\mathrm{Re}(n)k_\mathrm{A}R\bigr]
 e^{-2\mathrm{Im}(n)k_\mathrm{A}R}\Gamma_0,
\end{multline}
recall Eqs.~(\ref{e2.1}) and (\ref{e47}).
\begin{figure}[!t!]
\noindent
\includegraphics[width=1\linewidth]{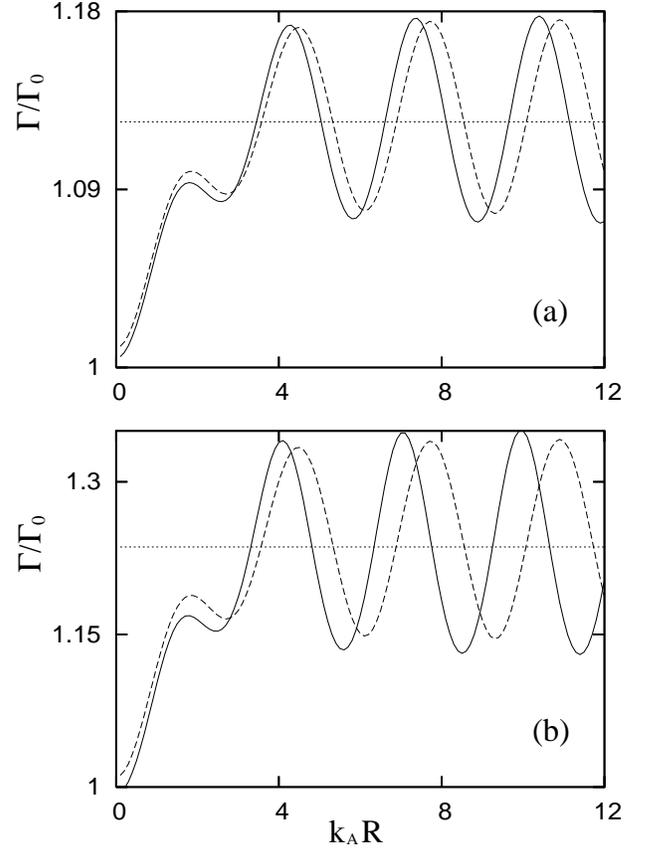}
\caption{
The local-field corrected spontaneous-decay rate (solid lines) and the
corresponding result in the linear Born approximation (dashed lines)
of an atom situated at the center of a dielectric sphere of
permittivitiy (a) $\varepsilon$ $\!=$ $\!1.1+10^{-8}i$ and (b)
$\varepsilon$ $\!=$ $\!1.2+10^{-8}i$ as a function of the sphere
radius $R$. The radius of the real (empty-space) cavity is
$R_\mathrm{C}$ $\!=$ $\!0.01/k_\mathrm{A}$. For comparison, the decay
rate in a bulk medium of the same permittivity (dotted lines) is also
plotted.
}
\label{cnt}
\end{figure}%
Comparison of Eq.~(\ref{e46}) with Eq.~(\ref{e2}) indicates that with
increasing size of the sphere, the validity of the linear Born
approximation to the decay rate becomes less satisfying in two
respects. (i) The oscillating term is not damped, and (ii) there is an
accumulated error in the phase. The former effect is insignificant
as long as $\mathrm{Im}(\chi)k_\mathrm{A}R$ $\!\ll$ $\!1$. This
restriction still allows for a large  range of sphere sizes; for
example, for $\mathrm{Im}\,\chi$ $\!\simeq$ $\!10^{-8}$ and
$\lambda_\mathrm{A}$ $\!=$ $\!2\pi/k_\mathrm{A}$ $\!\simeq$
$\!1\mu\mathrm{m}$, the condition \mbox{$R$ $\!\ll$
$\!10^7\mu\mathrm{m}$} follows.

{F}igure~\ref{cnt} shows the dependence on the sphere radius of
the local-field corrected spontaneous-decay rates according to
the exact equation (\ref{Eq.80}) and the linear Born expansion
[Eq.~(\ref{e24}) together with Eqs.~(\ref{e24.1}) and (\ref{e44.1})]
for two values of the permittivity. For the parameters used, we have
$\mathrm{Im}\,\chi/(k_\mathrm{A}R_\mathrm{C})^3$ $\!=$ $\!10^{-2}$,
so that the first two terms in the curly brackets in
Eq.~(\ref{Eq.80}), which arise from absorption, are negligibly small
compared to the last one. As expected from Eqs.~(\ref{e2}) and
(\ref{e46}), the decay rate oscillates with the sphere radius around
the bulk value. For small spheres, the linear Born expansion is seen
to be in good agreement with the exact result. It is further seen
that, as the sphere radius increases, an increasing phase shift
develops between the exact rate and the rate in the linear Born
approximation. A comparison of Figs.~\ref{cnt}(a) and (b) reveals that
this phase shift is larger for larger values $\mathrm{Re}\,\chi$.

\begin{figure}[!t!]
\noindent
\includegraphics[width=1\linewidth]{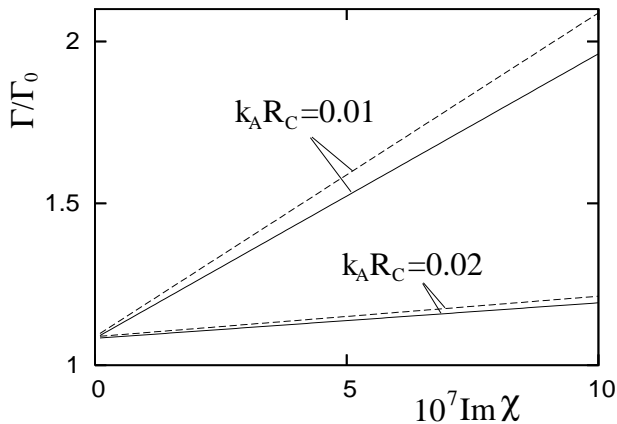}
\caption{
The local-field corrected spontaneous-decay rate (solid lines) and the
corresponding result in the linear Born approximation (dashed lines)
of an atom situated at the center of a dielectric sphere of radius $R$
$\!=$ $\!2/k_\mathrm{A}$ as a function of $\mathrm{Im}\,\chi$
($\mathrm{Re}\,\varepsilon$ $\!=$ $\!1.1$), for real-cavity radii
$R_\mathrm{C}$ $\!=$ $\!0.01/k_\mathrm{A}$ and $0.02/k_\mathrm{A}$.
}
\label{ci}
\end{figure}%
The dependence of the decay rate on the imaginary part of the
susceptibility is illustrated in Fig.~\ref{ci} for two values of
the real-cavity radius. It can be seen that the disagreement between the
curves representing the linear Born expansion and the exact result
increases with increasing material absorption or decreasing
cavity radius, i.e., with an increase in the combined factor
$\mathrm{Im}\,\chi/(k_\mathrm{A}R_\mathrm{C})^3$. In particular, if
in the case where $R_\mathrm{C}$ $\!=$ $\!0.01/k_\mathrm{A}$,
$\mathrm{Im}\,\chi$ changes from $10^{-8}$ to $10^{-6}$, then
$\mathrm{Im}\chi/(k_\mathrm{A}R_\mathrm{C})^3$ changes
from $10^{-2}$ to $1$, and the agreement worsens from being very good
to being moderately good. For the larger cavity radius $R_\mathrm{C}$
$\!=$ $\!0.02/k_\mathrm{A}$, which physically means a more dilute
medium, $\mathrm{Im}\chi/(k_\mathrm{A}R_\mathrm{C})^3$ varies from
$10^{-3}$ to $10^{-1}$ for the same variation of $\mathrm{Im}\,\chi$.
Throughout this range, the combined factor remains much smaller than
one and a good agreement is observed.

\subsubsection{Atom off center}
\label{subsec_offcnt}

Basing on a numerical computation of
$\Gamma_\mathrm{B}^{(1)\perp(\|)}$ as given by Eq.~(\ref{e26}) in the
linear Born approximation, we have also studied the case when the atom
is localized at an arbitrary position inside the sphere.
\begin{figure}[!t!]
\noindent
\includegraphics[width=1\linewidth]{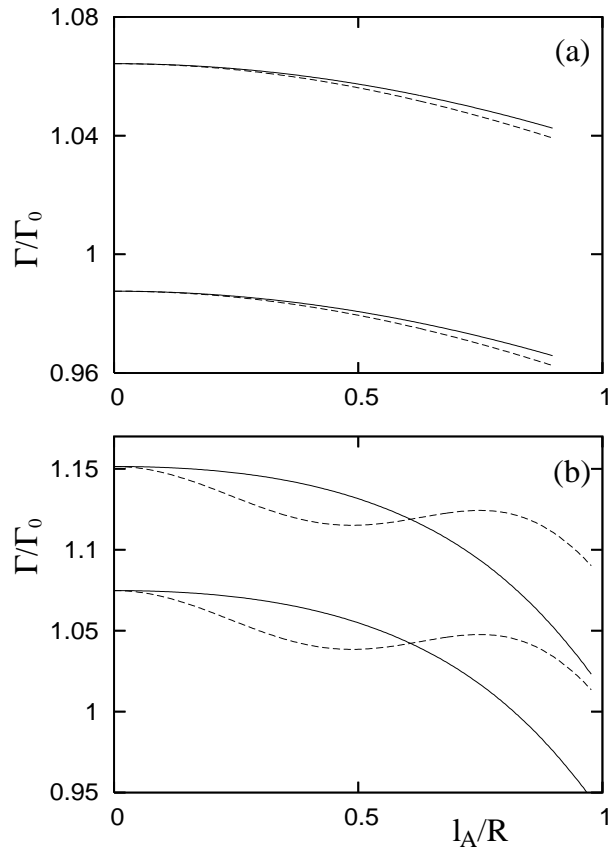}
\caption{
Position dependence of the local-field corrected (upper curves)
and uncorrected (lower curves) spontaneous-decay rate in the linear
Born approximation of an atom in a dielectric sphere of permittivity
$\varepsilon$ $\!=$ $\!1.1+10^{-8}i$ and radius \mbox{(a) $R$ $\!=$
$\!1/k_\mathrm{A}$} and (b) $R$ $\!=$ $\!5/k_\mathrm{A}$. The
real-cavity radius is $R_\mathrm{C}$ $\!=$ $\!0.01/k_\mathrm{A}$. The
solid (dashed) curves refer to radially (tangentially) oriented
transition dipole moments.
}
\label{offcnt}
\end{figure}%
{F}igure \ref{offcnt} compares the position dependences of
the local-field corrected decay rate, as given by Eq.~(\ref{e24})
together with Eqs.~(\ref{e24.1}) and (\ref{e26}), and the uncorrected
rate according to Eq.~(\ref{Eq.10}) (which is valid if absorption is
neglected), with the second term being given by Eq.~(\ref{e26}), for
two sphere radii (for the uncorrected rate beyond the linear Born
approximation, see also Ref.~\cite{Chew88}). From the figure it is
seen that when $l_\mathrm{A}$ $\!>$ $\!0$, i.e., when the atom is not
localized at the center of the sphere, radial and tangential
dipole orientations have to be distinguished, especially when the
sphere radius exceeds the wavelength of the radiation emitted by the
atom, such that interference effects begin to play a role. Note that
the whispering gallery resonances which may give rise to strong
enhancement near the rim of the sphere are not manifested here. The
existence of these resonances requires larger values of the real
part of the permittivity or larger sphere radii. Unfortunately, in
such cases the linear Born expansion is too rough an approximation to
the decay rate. From the figure it is further seen that for the
parameters used, the local-field correction increases the decay rate;
the amount of increase is determined essentially by the last term in
Eq.~(\ref{e24.1}).

\begin{figure}[!t!]
\noindent
\includegraphics[width=1\linewidth]{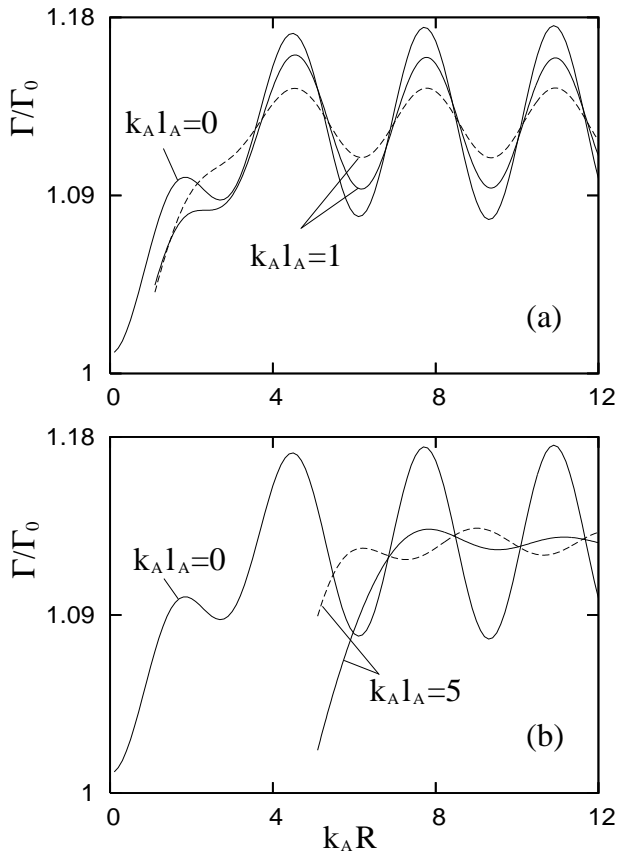}
\caption{
Decay rate of an atom located at (a) $l_\mathrm{A}$ $\!=$
$\!1/k_\mathrm{A}$ and (b) $l_\mathrm{A}$ $\!=$ $\!5//k_\mathrm{A}$
off center of a dielectric sphere of permittivity \mbox{$\varepsilon$
$\!=$ $\!1.1+10^{-8}i$} as a function of the sphere radius in linear
Born approximation. The cavity radius is $k_\mathrm{A}R_\mathrm{C}$
$\!=$ $\!0.01$. The solid (dashed) curves refer to radially
(tangentially) oriented transition dipole moments. For comparison, the
curves for an atom at the center \mbox{$k_\mathrm{A}l_\mathrm{A}$
$\!=$ $\!0$} are also shown.
}
\label{dp}
\end{figure}

In Fig.~\ref{dp} the dependence of the local-field corrected decay
rate on the sphere radius is illustrated. It can be seen that the
further the atom is displaced from the sphere center, the smaller the
amplitudes of oscillation of the decay rate become.

\section{Summary}
\label{sum}

Expressing the spontaneous-decay rate of an excited atom in the
presence of dielectric bodies in terms of the scattering part of the
associated Green tensor of the macroscopic Maxwell equations and
expanding the Green tensor in a Born series enables one to
systematically approach arbitrary geometries. In particular, in the
case of weakly polarizable bodies it may be possible to neglect
higher-order terms in the Born expansion and approximate the
scattering part of the Green tensor by the term linear in the
susceptibility, and the decay rate accordingly.

Using the real-cavity model of the local-field correction, we have
applied the theory to the problem of the local-field correction to the
spontaneous-decay rate of an excited atom embedded in a dispersing and
absorbing dielectric body of arbitrary size and shape. In this way, we
have derived a rate formula, which, within the linear Born
approximation, applies to atoms in arbitrary dielectric bodies, and
have given explicit conditions of its validity. To illustrate the
results, we have considered the case of an atom at an arbitrary
position inside spherical body in more detail.

It has surprisingly turned out that the rate formula found in linear
Born approximation agrees, to linear order in the susceptibility, with
a rate formula suggested by Toma\v{s} \cite{Tomas01} from his study
of the real-cavity model in the special and analytically solvable case
of a spherically symmetric system. We have then shown that this quite
general formula indeed applies to atoms in dielectric bodies of
arbitrary sizes and shapes, provided that the atoms are not in the
very vicinity of the surfaces of the bodies. So it can be shown that the
scattering part of the Green tensor that enters the basic formula for
the decay rate can always be decomposed into a term that only depends
on the properties of the local environment of the guest atom and the
local-field corrected scattering part of the Green tensor of the host
body without the guest atom, where the correction simply appears in
form of a factor. In particular, the former term can be substantially
determined by the absorptive properties of the host body, thereby
giving rise to a shift of the decay rate.

Finally, we note that in the same spirit as in the treatment of the
spontaneous decay, the Born expansion can also be employed in studying
other phenomena of the atom--field interaction in realistic systems
whose Green tensors are not known or analytically too involved.
Typical examples may be the Casimir-Polder interaction of a
ground-state atom with an inhomogeneous body \cite{Buhmann05},
the resonance fluorescence of an atom near such a body, or the
resonant energy transfer between atoms embedded in realistic bodies.


\acknowledgments

We would like to thank M.~S.~Toma\v{s} for discussions. H.T.D. thanks
the Alexander von Humboldt Stiftung and the National Program for Basic
Research of Vietnam for support. This work was supported by the
Deutsche Forschungsgemeinschaft.


\end{document}